\documentclass[aps,prl,a4,twocolumn,superscriptaddress,nofootinbib,preprintnumbers]{revtex4}

\usepackage{amsfonts}
\usepackage{mathrsfs}
\usepackage{amssymb}
\usepackage{amsmath}
\usepackage{graphicx,epsfig}
\usepackage{pstricks}
\usepackage{slashed}
\usepackage{dcolumn}%

\def\bs{\begin{small}}
\def\es{\end{small}}
\def\be{\begin{equation}}
\def\ee{\end{equation}}
\def\bea{\begin{eqnarray}}
\def\eea{\end{eqnarray}}
\def\bean{\begin{eqnarray*}}
\def\eean{\end{eqnarray*}}
\def\bary{\begin{array}}
\def\eary{\end{array}}
\def\bit{\begin{itemize}}
\def\eit{\end{itemize}}

\def\su5u1{SU(5) \times U(1)}
\def\fsu5u1{SU(5) \times U(1)'}
\def\so10{SO(10)}
\def\sq20{SO(10) \times SO(10)}

\def\bwt{\begin{widetext}}
\def\ewt{\end{widetext}}
\def\be{\begin{equation}}
\def\ee{\end{equation}}
\def\bea{\begin{eqnarray}}
\def\eea{\end{eqnarray}}
\def\bean{\begin{eqnarray*}}
\def\eean{\end{eqnarray*}}
\def\bary{\begin{array}}
\def\eary{\end{array}}
\def\bit{\begin{itemize}}
\def\eit{\end{itemize}}

\def\su5u1{SU(5) \times U(1)}
\def\fsu5u1{SU(5) \times U(1)'}
\def\so10{SO(10)}
\def\sq20{SO(10) \times SO(10)}

\usepackage{amssymb}

\begin{document}

\setlength{\parskip}{0.1cm}

\preprint{MIFPA-12-19,\, OSU-HEP-12-05}

\title{\large 
Unity of elementary particles and forces for the third family}

\author{Shreyashi Chakdar}
\email{chakdar@okstate.edu}
\affiliation{Department of Physics and Oklahoma Center for High
Energy Physics, Oklahoma State University, Stillwater, OK
74078-3072, USA}
\author{Tianjun Li}
\email{tli@itp.ac.cn}
\affiliation{ State Key Laboratory of Theoretical Physics 
and Kavli Institute for Theoretical Physics China (KITPC),
      Institute of Theoretical Physics, Chinese Academy of Sciences,
Beijing 100190, P. R. China }
\affiliation{George P. and Cynthia W. Mitchell Institute for
Fundamental Physics and Astronomy, Texas A$\&$M University,
College Station, TX 77843, USA}
\author{S. Nandi}
\email{s.nandi@okstate.edu}
\affiliation{Department of Physics and Oklahoma Center for High
Energy Physics, Oklahoma State University, Stillwater, OK
74078-3072, USA}
\author{Santosh Kumar Rai}
\email{santosh.rai@okstate.edu}
\affiliation{Department of Physics and Oklahoma Center for High
Energy Physics, Oklahoma State University, Stillwater, OK
74078-3072, USA}

\begin{abstract}

We propose a non-supersymmetric $SU(5)$ model in which only the third family of fermions are unified. The 
model remedies the non-unification of the three Standard Model couplings in non-supersymmetric $SU(5)$. 
It also provides a mechanism for baryon number violation which is needed for the baryon asymmetry of the 
Universe and is not present in the Standard Model. Current experimental constraints on the leptoquark gauge 
bosons, mediating such baryon and lepton violating interactions in our model, allow their masses to be at the 
TeV scale. These can be searched for as a ($b\tau$) or ($tt$) resonance at the Large Hadron Collider as 
predicted in our model.

\end{abstract}



\maketitle

\newpage


{\bf{Introduction}}: In the highly successful Standard Model (SM) of elementary particle interactions, 
the baryon and lepton numbers happen to be accidental global symmetries of the renormalizable 
interactions, and are conserved. However, there is no 
fundamental reason why these symmetries should be exact in nature. In fact, 
in the SM, the lepton number is violated by the SM gauge invariant 
dimension-five operators~\cite{Weinberg:1979sa}, while baryon number is violated 
by dimension-six operators~\cite{Weinberg:1979sa, Wilczek:1979hc}. The dimension-five operators 
can generate tiny neutrino masses~\cite{Weinberg:1979sa}, while the dimension-six operators can
cause proton decay~\cite{Weinberg:1979sa, Wilczek:1979hc}. 
However, if the ultra violet mass scale that suppresses these 
operators is the Planck scale, the generated neutrino masses are much smaller than 
the observed neutrino masses. Also, with the Planck scale suppression, the proton decay 
rate is too small to be observed in any future detector. The stability of 
proton was first questioned by Pati and Salam~\cite{PS},  and they proposed the
$SU(4)\times SU(2)_L \times SU(2)_R$ model (where the lepton number is the fourth
color)~\cite{PSLR} in which there are leptoquark gauge bosons violating both baryon 
and lepton numbers. These leptoquark gauge boson exchanges do not cause proton decay, but 
do cause $K_L \rightarrow {\mu} e $ transition~\cite{Valencia:1994cj}.
The current limit on this branching ratio, 
$B (K_L \rightarrow \mu e)= 4.7 \times 10^{-12}$ \cite{pdg} gives the mass limit on 
these leptoquark gauge bosons to be greater than $2300$ TeV. So this type 
of leptoquark gauge bosons is beyond the reach of the LHC. The minimal Grand 
Unified Theory (GUT) unifying the three SM gauge interactions was proposed by Georgi 
and Glashow with the $SU(5)$ gauge symmetry~\cite{Georgi:1974sy}.
However, the three SM gauge couplings do not unify in non-supersymmetric $SU(5)$.
 This model also has leptoquark 
gauge bosons $X_{\mu}$ and $Y_{\mu}$ leading to proton decay. 
Again the stability of the proton requires $M_{X_{\mu}},~M_{Y_{\mu}} > 10^{16}$~GeV. 
Same is true for the $SO(10)$ GUT~\cite{Fritzsch:1974nn}.

While the proton stability and $K_L \rightarrow {\mu} e $ process put severe limit on the 
masses of the leptoquark gauge bosons involving the first and second families of
the SM fermions, 
no such severe limit exists for the baryon and lepton number violating interactions involving 
the third family. For the first generation leptoquark searches at the 7 TeV LHC, CMS Collaboration 
with $36~ {\rm pb}^{-1}$ data has looked for the pair production of leptoquarks \cite{cms1}, 
and each decaying to $l q$ (with $l = e$ or $\nu$ and  $q$ being a light jet) with a branching 
ratio, $\beta = 1$ and $0.5$. They have set a limit, $M_{LQ} > 384$ GeV for the $eeqq$  
final state, and $M_{LQ} > 339$ GeV for the $e \nu q q $ final state. 
The corresponding 95\% C.L. limits on second generation leptoquarks from CMS with $2~ fb^{-1}$
of data is $M_{LQ} > 632(523)$  GeV for $\beta =1.0(0.5)$ \cite{cms2} while ATLAS with 
$1.03~ fb^{-1}$ of data set the limits as $M_{LQ} > 685(594)$  GeV for $\beta =1.0(0.5)$ \cite{atlas2}. 
For their third generation leptoquark search,  with $1.8~fb^{-1}$ of data in the final 
state $b b \nu \nu$, their limit is $M_{LQ} > 350$ GeV at $95\%$ C.L. \cite{cms3}. 
The bound from the Tevatron is weaker \cite{tevd0}. Thus for a leptoquark decaying to 
the third generation only, the limit on its mass is very low. In particular, 
a leptoquark decaying to $b \tau$ or $tt$ has not been looked at yet.

In this work, we propose a top $SU(5)$ model which remedies the non-unification  of the three SM couplings 
in the non-supersymmetric $SU(5)$ model. As our  non-supersymmetric  model is constructed using the 
$SU(5)\times SU(3)' \times SU(2)' \times U(1)'$, the SM couplings are combinations of the  corresponding 
couplings of ($g_5, g_3'$), ($g_5, g_2'$) and ($g_5, g_1'$); and thus no unification of the SM couplings  is needed. 
It also gives a mechanism for baryon number violation which is needed for the baryon asymmetry of the Universe 
and is not present in the SM.  The baryon and lepton violating gauge interactions involve only the third generation 
of the SM fermions. The leptoquark gauge bosons mediating these interactions are 
$(X_{\mu}, Y_{\mu})=({3,2,5/6})$ 
where the numbers inside the parenthesis represent the quantum numbers with 
respect to the SM $SU(3)_C\times SU(2)_L \times U(1)_Y$ gauge symmetry. 
The $X_\mu$ decays to $t t$ and $\bar{b} \tau^{+}$ while  $Y_\mu$ decays to $t b$, 
$\bar{t} \tau^{+} $, and $\bar{b} \nu$, which violate the baryon and lepton numbers. If we choose a basis such 
that the up-type quark mass matrix is  diagonal and the quark CKM mixings arise solely from 
the down-type quark sector, there will be no interactions of the  
$X u u$  or $Y u d$ type, thus preventing the proton decay. 
Since the leptoquark gauge bosons are color triplets, they can be pair produced 
($X\overline{X}$, $Y\overline{Y}$) if their masses are at the TeV scale. From their 
decays to $X \to {\bar{b} \tau^{+}}$ (or $\overline{X}\to{b \tau^{-}}$), one can reconstruct
the resonance by taking  suitable combinations of $b$ and $\tau$ 
in the final state, $ b b \tau^{+} \tau^{-}$. The same might be possible in the 
$tt$ channel if $t's$ can be reconstructed from their decay products. Below we 
present our model realizing this scenario.

{\bf{Top $SU(5)$ Model and the Formalism}}: Our model is an interesting unification 
of topcolor ~\cite{Hill:1991at,Hill:Dicus}, topflavor~\cite{Muller:Malkawi}, and top
hypercharge ~\cite{Chiang:2007sf} models. We call it top $SU(5)$ model. Our gauge symmetry is
$ SU(5)\times SM'$ where $ SM' = SU(3)'_C \times SU(2)'_L \times U(1)'_Y$. 
The first two families of the SM fermions are charged under $SM'$ and singlet under the $SU(5)$,
while the third family is charged under $SU(5)$ and singlet under  $SM'$.
We denote the gauge fields for $SU(5)$ and $SU(3)'_C \times SU(2)'_L \times U(1)'_Y $
as ${\widehat A}_{\mu}$ and ${\widetilde A}_{\mu}$, respectively,
and the gauge couplings as $g_5$, $g'_3$, $g'_2$ and $g'_Y$,
respectively. The Lie algebra indices for the generators of $SU(3)$, $SU(2)$ and $U(1)$
are denoted by $a3$, $a2$ and $a1$, respectively, and the  Lie algebra indices for the generators of
$SU(5)/(SU(3)\times SU(2)\times U(1))$ are denoted by ${\hat a}$.
After the $SU(5)\times SU(3)'_C \times SU(2)'_L \times U(1)'_Y $
gauge symmetry is broken down to the SM gauge symmetry $SU(3)_C\times SU(2)_L \times U(1)_Y$,
 we denote the massless gauge fields for the SM gauge symmetry as $A_{\mu}^{ai}$, and
the massive gauge fields as $B_{\mu}^{ai}$,
${\widehat X}_{\mu}^{\hat a}$ and ${\widehat Y}_{\mu}^{\hat a}$. The $X_{\mu}$ and 
$Y_{\mu}$ are the leptoquark gauge bosons. The gauge couplings for the SM gauge symmetry
$SU(3)_C$, $SU(2)_L$ and $U(1)_Y$ are $g_3$, $g_2$ and $g_Y$, respectively.

To break the $SU(5)\times  SM'$
gauge symmetry down to the SM gauge symmetry, we introduce
two bifundamental Higgs fields $U_T$ and $U_D$~\cite{Li:2004cj}. The fermion 
and Higgs field content of our models are shown in the first six rows of 
Table~\ref{tab:Content}. The first two family
quark doublets, right-handed up-type quarks, right-handed down-type 
quarks, lepton doublets, right-handed neutrinos, right-handed charged leptons,
and the corresponding Higgs field (belonging to $SM'$) are denoted 
as $Q_i$, $U_i^c$, $D_i^c$, $L_i$, $N^c_i$, $E_i^c$, and $H$ respectively. The third family 
SM fermions are $F_3$, $\overline{f}_3$, and $N_3^c$.
To give mass to the third family of the SM fermions, we introduce
an $SU(5)$ anti-fundamental Higgs field $\Phi\equiv (H'_T, H')$.
This would be the minimal top $SU(5)$ model in terms of field content. However, we then need 
to introduce the higher-dimensional (non-renormalizable) operators in the Higgs potential for the 
down-type quark Yukawa coupling terms between the first two families and third family. 
Instead,  we construct a renormalizable 
top $SU(5)$ model by introducing additional fields: the scalar field $XU$, 
and the vector-like fermions $(Xf,~\overline{Xf})$,
$(XD,~\overline{XD})$, and $(XL,~\overline{XL})$. 
To give the triplet Higgs $H'_T$ mass around 1 TeV, we also need to introduce
a scalar field $XT$. Otherwise, $H'_T$ will have mass around a few hundred
of GeV. The SM quantum numbers for these extra particles are given in Table~\ref{tab:Content} as well. 
 We shall present these two models in detail in a forthcoming paper~\cite{CLNR-P}. 
\begin{table}[htb]
\begin{center}
  \begin{tabular}{|c|c||c|c|}
   \hline
 Particles  & Quantum Numbers  & Particles & Quantum Numbers \\ \hline
 $Q_i$ & $({\bf 1}; {\bf {3}}, {\bf 2}, {\bf {1/6}})$  
& $L_i$ & $({\bf 1}; {\bf {1}}, {\bf 2}, {\bf {-1/2}})$  
\\\hline
 $U^c_i$ & $({\bf 1}; {\bf {\bar 3}}, {\bf 1}, {\bf {-2/3}})$  
& $N^c_k$ & $({\bf 1}; {\bf {1}}, {\bf 1}, {\bf {0}})$  
\\\hline
 $D^c_i$ & $({\bf 1}; {\bf {\bar 3}}, {\bf 1}, {\bf {1/3}})$  
& $E^c_i$ & $({\bf 1}; {\bf {1}}, {\bf 1}, {\bf {1}})$  
\\\hline
 $F_3$ & $({\bf 10}; {\bf {1}}, {\bf 1}, {\bf {0}})$  
& $\overline{f}_3$ & $({\bf {\bar 5}}; {\bf {1}}, {\bf 1}, {\bf {0}})$  
\\\hline
 $H$ & $({\bf 1}; {\bf {1}}, {\bf 2}, {\bf {-1/2}})$  
& $\Phi$ & $({\bf {\bar 5}}; {\bf {1}}, {\bf 1}, {\bf {0}})$  
\\\hline
 $U_T$ & $({\bf 5}; {\bf {\bar 3}}, {\bf 1}, {\bf {1/3}})$
& $U_D$ & $({\bf 5}; {\bf {1}}, {\bf { 2}}, {\bf -1/2})$
\\\hline \hline
$XT$ & $({\bf 1}; {\bf {\bar 3}}, {\bf 1}, {\bf {1/3}})$ &
$XU$ & $({\bf 10}; {\bf {1}}, {\bf 1}, {\bf {-1}})$ 
\\\hline 
 $Xf$ & $({\bf 5}; {\bf {1}}, {\bf 1}, {\bf {0}})$  
& $\overline{Xf}$ & $({\bf {\bar 5}}; {\bf {1}}, {\bf 1}, {\bf {0}})$  
\\\hline
$XD$ & $({\bf 1}; {\bf {3}}, {\bf 1}, {\bf {-1/3}})$  
& $\overline{XD}$ & $({\bf {1}}; {\bf {\bar 3}}, {\bf 1}, {\bf {1/3}})$  
\\\hline
$XL$ & $({\bf 1}; {\bf {1}}, {\bf 2}, {\bf {-1/2}})$  
& $\overline{XL}$ & $({\bf {1}}; {\bf {1}}, {\bf 2}, {\bf {1/2}})$  
\\\hline
\end{tabular}
\end{center}
\caption[]{The complete
particle content and the particle quantum numbers under 
$SU(5)\times SU(3)'_C \times SU(2)'_L \times U(1)'_Y $ gauge symmetry in the
top $SU(5)$ model. Here, $i=1,2$ and $k=1,2,3$.}
\label{tab:Content}
\end{table}

The Higgs potential breaking the $SU(5) \times SM'$ down to 
the SM gauge symmetry is given by
\begin{widetext}
\bs
\begin{eqnarray}
V = -m_{T}^2 |U_T^2|-m_{D}^2 |U_D^2| + \lambda_T |U_T^2|^2
+ \lambda_D |U_D^2|^2 + \lambda_{TD} |U_T^2| |U_D^2|
+\left[ A_T \Phi U_T XT^{\dagger} + A_D \Phi U_D H^{\dagger} +
 {{y_{TD}}\over {M_{*}}} U^3_T U^2_D + {\rm H.C.}\right]~,~\,
\label{potential}
\end{eqnarray} 
\es
\end{widetext}
The non-renormalizable $y_{TD}$ term is needed to give mass to the remaining Goldstone boson in our model, and 
is generated from the renormalizable interactions involving the
fields, $U_T$, $U_D$ and  $XU$, with $M_{*} \simeq M_{XU} \simeq 1000$ TeV ~\cite{CLNR-P}.

We choose the following vacuum expectation values (VEVs) for the fields $U_T$ and $U_D$
\bs
\begin{eqnarray}
 <U_T> =  {v_T} \left(
  \begin{array}{c}
    I_{3\times3} \\
    0_{2\times3} \\
  \end{array}
  \right)~, \quad
 <U_D> =  {v_D} \left(
  \begin{array}{c}
    0_{3\times2} \\
    I_{2\times2} \\
  \end{array}
  \right)~,
\end{eqnarray}\es
where $I_{i\times i}$ is the $i\times i$ identity matrix, and
$0_{i\times j}$ is the $i\times j$ matrix where all the entries are
zero. We assume that $v_D$ and $v_T$ are in the 
TeV range so that the massive gauge bosons  have TeV scale masses.

From the kinetic terms for the fields
$U_T$ and $U_D$ , we obtain the mass terms for the gauge fields 
\begin{small}
\begin{eqnarray}
&&\sum_{i=T, D} \langle (D_{\mu} U_i)^{\dagger} D^{\mu} U_i \rangle
=  {1\over 2} v_T^2   \left( g_5 {\widehat A}_{\mu}^{a3} - 
g'_3 {\widetilde A}_{\mu}^{a3} \right)^2 
 \nonumber \\ &&
+ {1\over 2} v_D^2  \left( g_5 {\widehat A}_{\mu}^{a2} - 
g'_2 {\widetilde A}_{\mu}^{a2} \right)^2
+\left( {{v_T^2}\over 3} 
+ {{v_D^2}\over 2}  \right) 
\left( g_5^Y {\widehat A}_{\mu}^{a1} - 
g'_Y {\widetilde A}_{\mu}^{a1} \right)^2
\nonumber\\ &&
+ {1\over 2} g_5^2 \left(v_T^2  +v_D^2 \right) 
\left(X_{\mu} \overline{X}_{\mu}
+ Y_{\mu} \overline{Y}_{\mu}\right)
~,~\,
\label{massterm}
\end{eqnarray}\end{small}
 where $g_5^Y \equiv {\sqrt 3} g_5/{\sqrt 5}$,
 and we define the complex fields
($X_{\mu}$, $Y_{\mu}$) and (${\overline{X}_{\mu}}$, ${\overline{Y}_{\mu}}$)
with quantum numbers (${ 3}$, ${ 2}$, ${ {5/6}}$) 
and (${ {\bar 3}}$, ${ 2}$, ${ -{5/6}}$), respectively
from the gauge fields ${\widehat A}_{\mu}^{\hat a}$, similar to
 the usual $SU(5)$ model~\cite{Georgi:1974sy}.

The $SU(5)\times SU(3)'_C \times SU(2)'_L \times U(1)'_Y $
gauge symmetry is  broken down to the diagonal SM gauge symmetry
$SU(3)_C\times SU(2)_L \times U(1)_Y$, and the theory is
unitary and renormalizable. The SM gauge couplings
$g_j$ ($j=3, 2$) and $g_Y$ are given by
\bs
\begin{eqnarray}
{1\over {g_j^2}} ~=~ {1\over {g_5^2}} + {1\over {(g'_j)^2}}~,~
{1\over {g_Y^2}} ~=~ {1\over {(g^Y_5)^2}} + {1\over {(g'_Y)^2}}~.~\,
\end{eqnarray}
\es

The renormalizable SM fermion Yukawa couplings are
\begin{small}\begin{eqnarray}
-{\cal L} &=& y^u_{ij} U_i^c Q_j {\widetilde H} +
y^{\nu}_{kj} N_k^c L_j  {\widetilde H} + 
 y^d_{ij} D_i^c Q_j H 
\nonumber\\ && 
+ y^e_{ij} E_i^c L_j H 
+y^u_{33} F_3 F_3 \Phi^{\dagger}
+ y^{d,e}_{33} F_3 {\overline f}_3 \Phi
\nonumber\\ &&
+ y^{\nu}_{k3} N_{k}^c  {\overline f}_3 \Phi^{\dagger}
+ m^N_{kl}  N_{k}^c  N_{l}^c  + {\rm H.C.},~~
\label{Yukawa-SM}
\end{eqnarray}\end{small}
where $i/j=1,~2$, $k/l=1,~2,~3$, and $ {\widetilde H}=i\sigma_2 H^{*}$
with $\sigma_2$ being the second Pauli matrix.
Because the three right-handed neutrinos can mix among
themselves via the Majorana masses, we can generate
the observed neutrino masses and mixings via TeV scale seesaw. 
In addition, the Yukawa terms between 
the triplet Higgs field $H'_T$ in $\Phi$ and
the third family of the SM fermions are
$ y^{d,e}_{33} t^c b^c H'_T$, $ y^{d,e}_{33} Q_3 L_3 H'_T$, 
and $y^{u}_{33} t^c \tau^c H^{\prime \dagger}_T$. So, we have
$(B+L)$ violating interactions as well.

It is worth pointing out here that we have chosen a basis in which the up-type quark Yukawa interactions and 
hence the up-quark mass matrix is diagonal. Therefore the quark CKM mixings need to be generated from the 
down-type quark sector. But the Yukawa couplings of Eq.~(\ref{Yukawa-SM}) have no mixing of the first and 
second families with the third family. So to generate the quark CKM mixing, we consider the 
dimension-five operators given by
\begin{small}
\begin{eqnarray}
-{\cal L} &=& {1\over {M_*}}
\left(  y^d_{i3} D_i^c F_3 \Phi U_T^{\dagger}
+ y^{e}_{i3} E_i^c {\overline f}_3 H U_D  
\right. \nonumber \\ && \left.
+ y^d_{3i} {\overline f}_3  Q_i H U_T
+ y^{e}_{3i} F_3 L_i \Phi U_D^{\dagger} \right) + {\rm H.C.} 
\label{O-Dim-5}
\end{eqnarray}\end{small}
The correct CKM mixings can be generated with $M_* \simeq 1000$ TeV. 
The dimension-five terms in Eq.~(\ref{O-Dim-5}) can be generated at 
the renormalizable level by using  
the vector-like fermions $(Xf,~\overline{Xf})$,
$(XD,~\overline{XD})$, and $(XL,~\overline{XL})$ with masses 
around 1000 TeV~\cite{CLNR-P}. Note that the dimension-five operators are generated using the 
vector like fermions only for the down sector. No such terms are generated for the up sector at the tree level
or radiatively.

We note that there is no proton decay in our model, because no up-type quark mixings
can be generated after we integrate out the vector-like particles. This can be seen as follows. The $SU(3)'_C\times SU(2)'_L \times U(1)'_Y$ gauge symmetry
can be formally embedded into a global $SU(5)'$ symmetry. Under
$SU(5)\times SU(5)'$, the bi-fundamental fields $U_T$ and $U_D$ form
$({\bf 5}, {\bf {\bar 5}})$ representation, the
vector-like particles $Xf$ and $\overline{Xf}$ respectively
form $({\bf 5}, {\bf { 1}})$ and $({\bf {\bar 5}}, {\bf 1})$ representations,
and the vector-like particles ($XD,~\overline{XL}$) and
($\overline{XD}, ~XL$)  respectively form $({\bf { 1}}, {\bf 5})$ and 
$({\bf { 1}}, {\bf {\bar 5}})$ representations. Because all these
fields are in the fundamental and/or anti-fundamental representations of $SU(5)$
and/or $SU(5)'$,  we cannot create the Yukawa interactions 
$10_f 10'_f 5_{H}$ or $10_f 10'_f 5_{H'}$ for the up-type quarks
after we integrate out the vector-like particles. Thus, there is no proton decay problem. 
We also note that there is no Landau pole in our model. The ultraviolet cutoff scale could 
be the Planck scale, since the $SU(5)$ is asymptotically free.

{\bf{Phenomenology and LHC Signals}}: Leptoquark production at LHC will have large 
cross sections \cite{blumlein:belyaev}. The leptoquark gauge bosons, $X_{\mu}$ and $Y_{\mu}$ 
can be pair produced at the LHC, {\it viz.} $pp\rightarrow X \overline{X}$ and $pp\rightarrow Y \overline{Y}$.  
In our model, the decay modes of $X_{\mu}$ are to $\bar{b} \tau^{+}$ and $tt$, with the 
former mode dominating at the low $X_{\mu}$ mass region. The modes of $Y_{\mu}$ are to $\bar{b} \nu$, 
$\bar{t} \tau^+$, and $t b$. 

We consider here, the signal at the 7 TeV and 8 TeV runs of LHC, coming from the 
$X\overline{X}$ production, with $X_{\mu}$ decaying to $\bar{b}\tau^{+}$, and $\overline{X}_{\mu}$ 
decaying to $b \tau^{-}$, as these will be relatively less difficult 
modes to reconstruct the mass of $X_{\mu}$ and $\overline{X}_{\mu}$ from the decays. 
The final state signal is $b \bar{b} \tau^{+} \tau^{-}$ with all four particles being detected 
in the flavor tagged mode, {\it albeit} with respective tagging efficiencies. Although the 
$\tau$ modes can be distinguished by their charge, the $b$ and $\bar{b}$ cannot be 
distinguished from each other. Thus to reconstruct the mass of the $X_{\mu}$ we need to pair the
the $\tau^\pm$ with both the $b$ jets. The dominant SM background for our final state
comes from $pp\to 2b2\tau,4b,2j2b,2j2\tau,4j,t\bar{t}$ where $j=u,d,s,c$.  The light jet final 
states can be mistagged as $\tau$ or $b$ jets and thus form a significant source for the background
due to the large cross sections at LHC, as they are dominantly produced through strong interactions. 
We find that at leading-order and with a kinematic selection of $p_T>15$ GeV, $|\eta|<2.5$
and $\Delta R>0.2$  for all four particles, the SM background at LHC with $\sqrt{s}=7$ TeV is 
$\simeq 9.7$ pb while it is $\simeq 11.7$ pb at LHC with $\sqrt{s}=8$ TeV estimated 
using {\tt Madgraph 5} \cite{mad5}. However the background is 
significantly suppressed to $\sim 1.6$ fb when we choose stronger cuts of $p_T>80$ GeV and 
$\Delta R> 0.4$ for all particles. Also for similar cuts the SM background at LHC with 
$\sqrt{s}=8$ TeV is estimated to be around $\sim 2.5$ fb. 
Note that we have used the following efficiencies for $b$ and $\tau$ tagging, 
$\epsilon_b=\epsilon_\tau=0.5$ while we assume a mistag rate for light jets to be 
tagged as $b$ or $\tau$ as 1\% and $c$ jets tagged as $b$ jets to be 10\%.     
For analyzing the signal we choose two values for the mass of $X_\mu$, {\it viz.} 
$M_X=600 (800)$ GeV which are pair produced at 7 and 8 TeV run of LHC with production cross sections 
of $\sim 275.5 (559.5)$ fb and $\sim 23.5 (55.7)$ fb respectively. The pair produced leptoquarks 
would then decay to give us the $b \bar{b} \tau^{+} \tau^{-}$ final state. To account for the 
detector resolutions for energy measurement of particles, we have used a Gaussian 
smearing of the jet and $\tau$ energies with an energy resolution given by $\Delta E/E = 0.8/\sqrt{E~(GeV)}$ 
and $\Delta E/E = 0.15/\sqrt{E~(GeV)}$ respectively when analyzing the signal events. The strong cuts on the 
final states do not affect the signal too much as the final state particles come from the decay of a heavy parent 
particle and therefore carry large transverse momenta. This gives us cross sections for the $2b2\tau$ final state 
as 7.5 fb for $M_X=600$ GeV and 0.62 fb for $M_X=800$ GeV which were 8.4 fb and 0.67 fb for the two masses 
respectively, with the less stringent cuts at LHC with $\sqrt{s}=7$ TeV. Similarly, one finds that for the current 
run of LHC with $\sqrt{s}=8$ TeV, we get signal cross sections of 14.8 fb for  $M_X=600$ GeV and 1.5 fb for 
$M_X=800$ GeV with the stronger kinematic cuts. The corresponding numbers with the less 
stringent cuts were 17 fb and 1.6 fb respectively. Note that we have included the tagging 
efficiencies and the corresponding branching fraction of the $X_\mu$ decaying to the $b\tau$ 
mode in evaluating the above quoted numbers for the signal cross section. A quick look at the 
signal and SM background cross sections shows that a resonance in the invariant mass distribution of the 
$b\tau$ final state for the signal for mass $M_X=600$ GeV at LHC with  $\sqrt{s}=7$ TeV
and for $M_X=800$ GeV 
at LHC with $\sqrt{s}=8$ TeV would clearly stand out against the very small SM background. 
\begin{widetext}
\begin{center}
\begin{figure}[!ht]
\includegraphics[width=2.8in,height=2.6in]{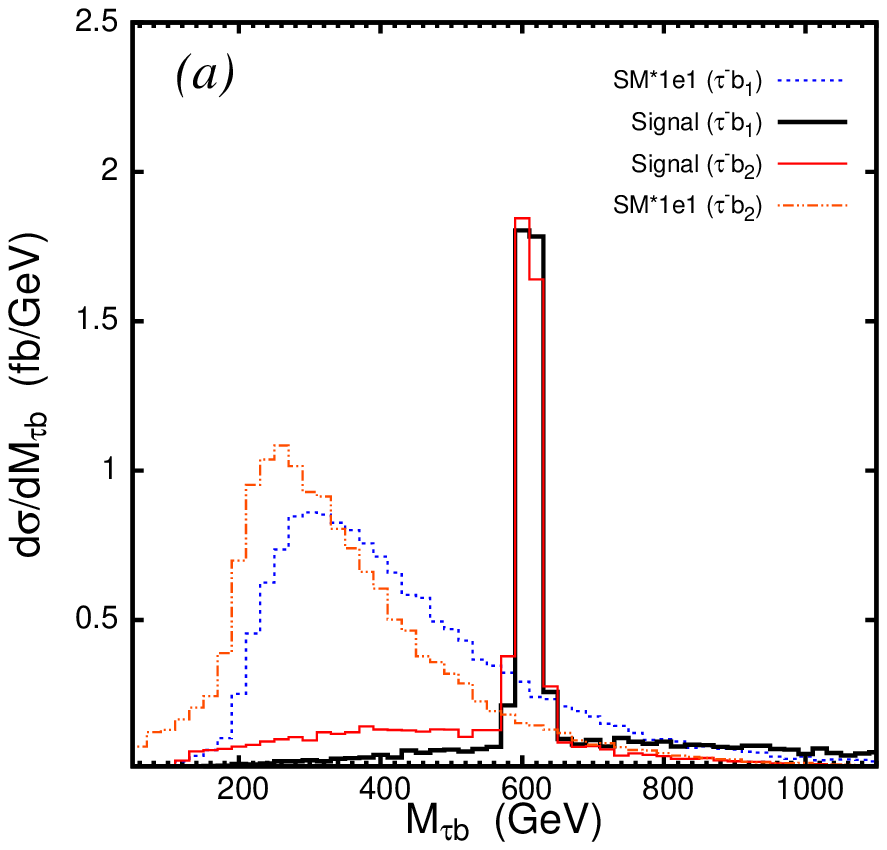}
\includegraphics[width=3.1in,height=2.6in]{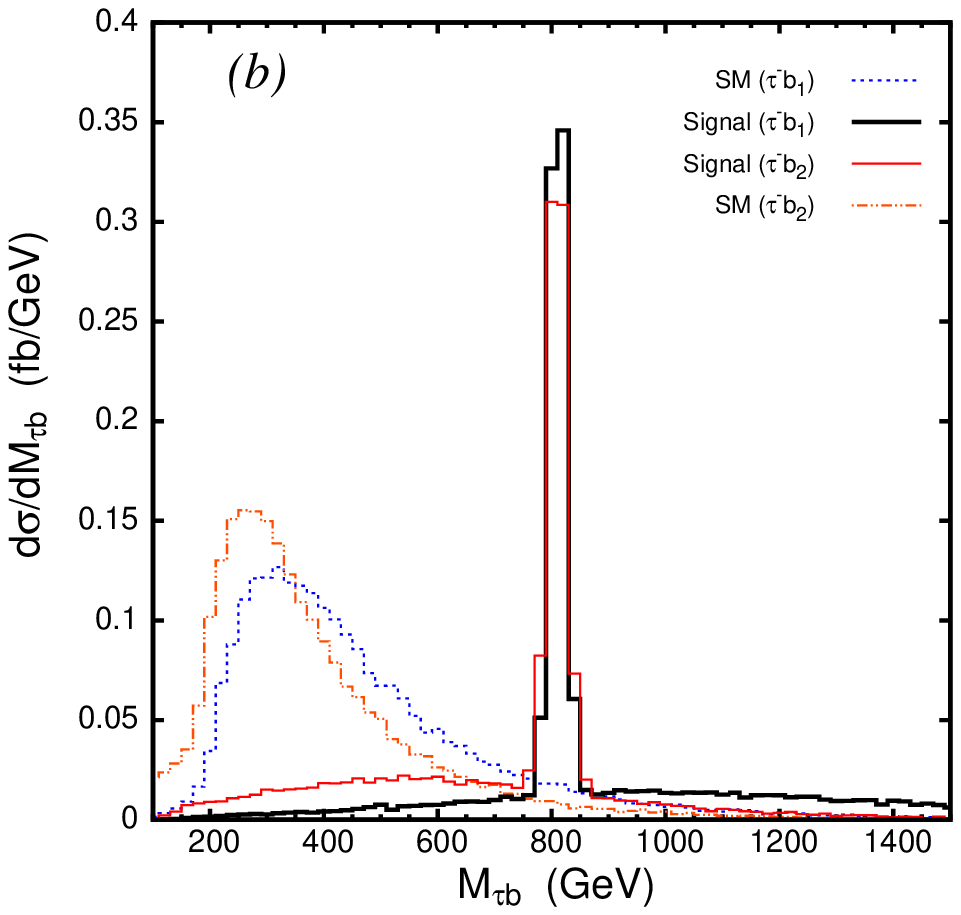}
\caption{The invariant mass distribution in $\tau^-b$ for the signal 
corresponding to 
(a) $M_X=600$ GeV at LHC with $\sqrt{s}=7$ TeV and (b) $M_X=800$ GeV at 
LHC with $\sqrt{s}=8$ TeV. Also included are the SM 
distributions where the background has been multiplied by a factor of 
$10$ in (a).}
\label{Minv.fig}
\end{figure}
\end{center}
\end{widetext}

To put this in perspective, in Fig.\ref{Minv.fig}, we plot the invariant mass distribution of the $\tau^-$ with either of the 
$b$ jets which are ordered according to their $p_T$ for the two choices of the $X_\mu$ mass.    
We clearly see the leptoquark ($X_\mu$) peak around 600~GeV and 800~GeV in the signal while the SM events 
fall off rapidly at high values of the invariant mass. Note that as the $b$ jets are ordered according to their $p_T$, so either 
can form the correct combination with the charged $\tau$ for the peak and thus both distributions lead to a 
peak in the invariant mass. It is also worth noting that if such a $p_T$ ordering is used then 
either $b$ jet combined with either of the charged $\tau$ will give an invariant mass peak 
at the same mass. The SM background is quite suppressed compared to the signal for LHC with  $\sqrt{s}=7$ TeV 
and is shown after multiplying by a factor of $10$ in Fig.\ref{Minv.fig}$(a)$.  As can be seen from Fig.\ref{Minv.fig}, the 
signals are clearly visible above the background. Therefore a dedicated search in invariant mass bins in the $b\tau$ channel 
will be very useful in searching for such a leptoquark signal, even with small signal cross sections.  To highlight this we 
also estimate that with the data available (5 $fb^{-1}$) at 7 TeV collisions at LHC, the leptoquark in our model will give 5 
signal events for mass as high as 750 GeV. The reach would be further improved at the current run of LHC with center 
of mass energy of 8 TeV.  We find that we get 5 signal events with data corresponding to an integrated luminosity of 
5 $fb^{-1}$  already collected at 8 TeV collisions, for leptoquark mass of 840 GeV while with an integrated luminosity 
of 15 $fb^{-1}$ achievable in the near foreseeable future one can get 5 signal events for leptoquark mass as high as 
940 GeV.  Another promising final state is $t t b \tau^{-}$ arising from the decays $X\to tt$, and 
$\overline{X}\to \bar{b} \tau^{-}$, if the top quarks can be reconstructed. The pair production of the $Y_\mu$ leptoquark 
gauge bosons also lead to many interesting signals. Details of these and other multijet and multilepton final states with or 
without missing energy will be discussed in a forthcoming publication~\cite{CLNR-P}.

{\bf Summary and Conclusions:}
We have presented a top $SU(5)$ model which remedies the non-unification of the three SM couplings in the 
non-supersymmetric $SU(5)$ model. The model has baryon and lepton number violation which is needed to explain 
the baryon asymmetry of the Universe and is not present in the SM.
Our model is renormalizable and  satisfy all the 
existing experimental constraints, and do not cause proton decay. The gauge bosons, $X_{\mu}$ and $Y_{\mu}$, which 
mediate baryon and lepton number violating interactions, involve only the third family of fermions, and 
 can be pair produced at the LHC. 
$X_{\mu}$ can be reconstructed as a $\bar{b} \tau^{+}$ resonance in the four jet final state, as well as, possibly 
in the ($tt$) mode. We encourage our ATLAS and CMS colleagues to search for these leptoquark 
gauge bosons in the proposed final states.  

\begin{acknowledgments}

We thanks A. Khanov of the ATLAS Collaboration and R. Rahmat of the CMS Collaboration for
useful discussions. This research was supported in part by the Natural Science Foundation of China under grant numbers 
10821504, 11075194, and 11135003, 
and by the United States Department of Energy Grant Numbers DE-FG03-95-Er-40917 and  DE-FG02-04ER41306.

\end{acknowledgments}


\end{document}